\documentstyle[12pt,epsfig]{article}
\begin{document}
\begin{center}
{\Large {\bf Motion of charged particles around a rotating black hole
in a magnetic field}} \\[5mm]
{\large A. N. Aliev $^{\dagger}$  and  N. \" Ozdemir $^{\dagger\ddagger}$}
\\[2mm]
$^{\dagger}$  Feza G\"ursey Institute, P.K. 6  \c Cengelk\" oy,
81220 Istanbul, Turkey \\
$^{\ddagger}$ ITU, Faculty of Sciences and Letters, Department of Physics,\\
80626 Maslak, Istanbul, Turkey \\ [5mm] \today
\end{center}
%\noindent

\footnote{email address: aliev@gursey.gov.tr}

%\newpage
\begin{center}
{\bf ABSTRACT}
\end{center}
\noindent
We study the effects of an external magnetic field, which is assumed
to be uniform at infinity, on the marginally stable
circular motion of charged particles in the equatorial plane of
a rotating black hole. We show that the magnetic field has
its greatest effect in enlarging the region of stability
towards the event horizon of the black hole.
Using the Hamilton-Jacobi formalism
we obtain the basic equations governing the marginal stability
of the circular orbits and their associated energies and angular momenta.
As instructive examples, we review the case of the marginal
stability of the circular orbits in the Kerr metric,
as well as around a Schwarzschild black hole in a magnetic field.
For large enough values of the magnetic field around a maximally
rotating black hole we find the limiting analytical solutions
to the equations governing the radii of marginal stability.
We also show that the presence of a strong magnetic field
provides the possibility of relativistic motions in
both direct and retrograde innermost stable circular orbits
around a Kerr black hole.

\vspace{4mm}

\newpage

\section{Introduction}

New observational evidence for black holes provides new motivations
for the investigation of the general relativistic dynamics of
particles and electromagnetic fields in the vicinity of the black holes.
We shall start with a brief description of the situation.
The results of astronomical observations over the last decade
continue to point insistently to the existence of
stellar-mass and supermassive black holes
in some X-ray binary systems and in galactic centres
(see  Horowitz $\&$  Teukolsky 1999; Rees 1998 for reviews).
The typical examples of the stellar-mass black holes
in X-ray binaries are Cyg X-1 discovered back in 1971, the
X-ray source LMC X-3 in the Large Magellanic Cloud, as well as
the source in A0620-00  discovered in 1975
and a number of recently discovered sources, such as V404 Cyg,
GS 2000+25, GRO J0422+32 (for full list see Charles 1999).
New observational data, such as the detection of broad iron
fluorescence lines and maser emission lines of water in the spectra
provide the strongest suggestion for the
presence of supermassive black holes in the centres of
the active galaxies MCG 6-30-15, NGC 4258 and NGC 1068
Menou, Quataert  $\&$ Narayan 1999; Rees 1998; Miyoshi et.al 1995;
Watson $\&$ Wallin 1994).

The supermassive black holes are also strongly believed
to be in the centres of some
low-level active, or non-active, galaxies. For instance,
recent progress in the studies of the
distributions and velocities of stars near the centres of
the giant elliptical galaxy M87, Andromeda M31 and our
own Galaxy have revealed the strong evidence
for the existence of the supermassive black holes in these centres
(Merritt $\&$ Oh 1997; Richstone et al. 1998; Rees 1982, 1998).

On the other hand, a convincing explanation for a
huge amount of energy output from the active cores of the galaxies
associated with the supermassive black holes requires the searches
for mechanisms responsible for the high-level energy release.
One of these mechanisms is the extraction of the rotational energy
from a rotating black hole surrounded by stationary magnetic fields.
The magnetic fields threading the event horizon tap the rotational energy
of the black hole due to the interaction between charged particles
and the induced electric field (Blandford $\&$ Znajek 1977;
Thorne, Price $\&$ Macdonald 1986). The interest in this model
still continues to point out new gravito-electromagnetic phenomena
in the strong and weak field domains around a rotating black hole
(Bi\u c\'ak $\&$ Ledvinka 2000; Chamblin, Emparan $\&$ Gibbons 1998;
Mashhoon 2001; see also Punsly 2001). Another mechanism
responsible for high-level energy release
is gas accretion by a black hole (see Shapiro $\&$ Teukolsky 1983), where
energy is released mostly at the expense of the binding energy of the
particles and the strong gravitational field of the black hole.
It is well known that the binding energy in the innermost stable orbits
of the particles determines the potential efficiency of an
accretion disc. It is about 5.7 \% of the rest energy in the
Schwarzschild field, but in the case of a maximally rotating black hole
it approaches 42 \% of the rest energy. The observational data
from the core of some galaxies, such as the elliptical galaxy M87 reveal
that the inner part of a gas disc
around a putative supermassive black hole emits non-thermal
radiation and radio waves. The reason for this is believed
to be synchrotron emission from ultrarelativistic electrons
moving in a strong magnetic field in the inner
part of the accretion flow onto the supermassive black hole
(Fabian \& Rees 1995; Narayan \& Yi 1995; Rees 1998).
This gives us a new impetus to return back once again to the
investigation of the motion of charged particles in the
model of a rotating black hole in a uniform magnetic field,
though much insight into this problem was given in 1980s
(Prasanna $\&$ Vishveshwara 1978; Prasanna 1980;
Wagh, Dhurandeur \& Dadhich 1985; Aliev $\&$ Gal'tsov 1989;
see also Frolov $\&$ Novikov 1998 and references therein).
In particular, the works of Prasanna \& Vishveshwara (1978) and
Prasanna (1980) have given a comprehensive numerical analysis of
the charged particle motion in a magnetic field superposed on
the Kerr metric by studying the structure
of the effective potential for radial motion and integrating
the complete set of equations of the motion for appropriate
initial conditions.

In this paper we shall study a particular class of
orbits, namely the marginally stable circular orbits
of charged particles in the equatorial plane of a
Kerr black hole immersed in a uniform magnetic field.
The numerical analysis performed
by Prasanna \& Vishveshwara (1978) has revealed that the presence
of a uniform magnetic field on the Kerr background
increases the range of stable circular orbits. This result
is confirmed in our model of the marginally stable circular
motion, however we also obtain significant new results:
First of all, we derive the basic equation
determining the region of the marginal stability of the circular orbits,
that comprises only two parameters, namely the rotation parameter
of black hole and the strength of the magnetic field.
We also find the closed analytical expressions for the associated angular
momentum and energy of a charged particle moving in a marginally
stable circular orbit. Further, for a sufficiently large values of
the magnetic field, as well as for a maximum value of the black hole
rotation parameter we find the limiting analytical solutions 
for the radii of stability of both direct and
retrograde innermost circular orbits. This allows us not only 
to emphasize that the presence of a magnetic field enlarges
the region of stability towards the event horizon, but also
to find the limiting values for this enlargement both for direct
and retrograde motions along with their associated energies
and angular momenta.
Our analytical and numerical calculations show that the combined
effects of a sufficiently strong magnetic field and the rotation of a
black hole give rise to the possibility of relativistic motion of
the charged particles in the innermost stable direct
and retrograde orbits. The existence of relativistic motion
in the innermost stable direct orbits is especially
important new feature worked out in our analysis.
These orbits lie very close to the event horizon
of a rotating black hole and they may provide a mechanism
for synchrotron emission from relativistic charged particles
with the signatures of the strong-gravity domain.

The paper is organized as follows. We shall first review
the solution of the Maxwell equations describing
a uniform magnetic field in the Kerr metric
with a small electric charge (see Section 2). In Section 3 we consider
the separation of variables in the Hamilton-Jacobi equation and derive
the effective potential for the radial motion of charged particles
around a Kerr black hole in a uniform magnetic field. These results are
used in Section 4 to obtain the basic equations governing
the region of the marginal stability of the circular orbits and their
associated energies and angular momenta. To make the analysis more
transparent in the general case, we shall first present
the results of analytical and numerical
calculations for the marginal stability of the circular orbits
in the pure Kerr  metric, as well as in a uniform magnetic field in the
Schwarzchild metric (see Sections 4.1 and 4.2). Section 4.3 is devoted to
a comprehensive analysis of the effect of a uniform magnetic field
on the radii and the assigned energy and angular momentum of the
marginally stable circular orbits for various values of the magnetic field
and the rotation parameter of the black hole.

\begin{center}
\section{Uniform magnetic field around a rotating black hole}
\end{center}

It is well known that an electrically neutral black hole can not have
intrinsic magnetic field (Ginzburg $\&$ Ozernoy 1965). However,
a magnetic field near a black hole can arise due to external factors,
such as the presence of a nearby magnetars or neutron stars. Accretion
of a matter cloud may also form a magnetosphere with a superstrong
magnetic field of magnitude up to
$ B_M={\textstyle\frac{1}{M}}
\simeq 2\times 10^{10} (M/10^9 M_{\odot})^{-1} G$
around a supermassive black hole (Kardashev 1995).
This value corresponds to the case when the magnetic and
gravitational pressures become equal to each other
(Bi\u c\'ak $\&$ Jani\u s 1985).
It is clear that with this value of the magnetic field
the space-time geometry near a black hole will be significantly distorted
(Aliev $\&$ Gal'tsov 1989), however, when $ B \ll B_M $
there is definitly a region near the black hole where the space-time
is not distorted by the external magnetic field and the latter can be
considered as a perturbation. We shall consider this case i.e.
a rotating black hole with a small electric charge $(Q \ll M) $ immersed
in an external magnetic field described by a corresponding solution
of the Maxwell equations against the background of the Kerr metric
\begin{eqnarray}
ds^2 & = &\left(1-\frac{2 M r}{\Sigma} \right) dt^2+
\frac{4 M a r}{\Sigma} \sin^2 \theta\,dt \,d\phi
-\frac{A \sin^2\theta}{\Sigma}\,d\phi^2  \nonumber \\
&  &-\Sigma\,\left(\frac{dr^2}{\Delta} + d\theta^2 \right) ,
\label{metric}
\end{eqnarray}
where $M$ is the mass of the black hole, $a=J/M$ is its angular momentum
per unit mass, $ A=(r^2+a^2)^2-\Delta a^2 \sin^2\theta$,
$\Delta=r^2+a^2-2 M r $ and $\Sigma=r^2+a^2 \cos^2 \theta $. We shall
assume a magnetic field around the Kerr black hole
to be uniform at infinity. In this case there is an elegant way
to construct the solution of the Maxwell equation (Wald 1974).
We shall now give a succinct description of this solution.

The Kerr space-time is stationary and axially symmetric that implies
the existence of two commuting Killing vector fields
$\xi^{\mu}_{(t)}=(1,0,0,0)$ and $\xi^{\mu}_{(\phi)}=(0,0,0,1)$.
These fields are used as a $4$-vector potential $A_{\mu}$
describing electromagnetic fields in the Kerr metric
that are superpositions of Coulomb electric and
asymptotically uniform magnetic fields.
Indeed, for Ricci-flat space-times $(R_{\mu\nu}=0)$,
the Maxwell equations for the $4$-vector potential in the covariant
Lorentz gauge
$A^{\mu}_{\;;\;\mu}=0 $
\begin{equation}
{A^{\mu}_{\;;\;\nu}}^{;\;\nu}=0
\label{max1}
\end{equation}
and the equation
\begin{equation}
{\xi^{\mu}_{\;;\;\nu}}^{;\;\nu} =0
\label{kil2}
\end{equation}
for a Killing vector are the same. The semicolon means covariant
differentiation. We shall take the $4$-vector potential
in the form
\begin{equation}
A^{\mu} = \alpha\, \xi^{\mu}_{(t)}+ \beta \, \xi^{\mu}_{(\phi)}
\label{pot1}
\end{equation}
where $\,\alpha\,$ and $\,\beta\,$ are arbitrary parameters.
Next, we calculate the Maxwell $2$-form using
equation (\ref{pot1}). We find
\begin{eqnarray}
F& =& \frac{2 M}{\Sigma}\left(1-\frac{2 r^2}{\Sigma}\right) (\alpha- \beta a
\sin^2\theta)(dt\wedge dr +a \sin^2\theta \,dr\wedge d\phi) \nonumber \\
& &
-\,\beta \,(2 r \sin^2\theta \,dr\wedge d\phi + \frac{A}{\Sigma} \sin 2\theta\,
d\theta \wedge d\phi) \nonumber \\ [2mm]
& &
+\,\frac{2 M a r}{\Sigma^2} \sin^2\theta\, [(\alpha a -\beta (r^2+a^2))\,
dt\wedge d\theta \nonumber \\ [2mm]
& &
+\,(r^2+a^2)(\alpha -\beta a \sin^2 \theta)\,
d\theta \wedge d \phi] \;,
\label{2form}
\end{eqnarray}
which in the asymptotic region $r \gg M $ reduces to
\begin{equation}
F = -2 \beta r \sin\theta \,(\sin\theta \,dr \wedge d\phi+ r \cos\theta\,
d\theta \wedge d\phi).
\label{2form1}
\end{equation}
It follows that $ \,\beta=B/2 \,$, where $ B $ is the strength of
the uniform magnetic field, which is parallel to the rotation axis
of the black hole. As for the remaining parameter $ \,\alpha\, $ it can be
specified using the surface integrals for the mass and angular momentum
of the black hole (Bardeen, Carter $\&$ Hawking 1973)
\begin{equation}
M = \frac{1}{8\pi} \int {\xi^{\mu}_{(t)}}^{;\;\nu} d^2\Sigma_{\mu\nu}
\;\;\;\;\;\;\;\;
J = -\frac{1}{16\pi} \int {\xi^{\mu}_{(\phi)}}^{;\;\nu} d^2\Sigma_{\mu\nu}
\label{mass}
\end{equation}
along with the Gauss integral for its electric charge
\begin{equation}
Q = \frac{1}{4\pi} \int F^{\mu \nu} d^2\Sigma_{\mu\nu} \,.
\label{gauss}
\end{equation}
Finally, we obtain
\begin{equation}
\alpha=a B - \frac{Q}{2M}
\label{alpha}
\end{equation}
Thus, the $4$-vector potential takes the form
\begin{equation}
A^{\mu} = \frac{1}{2} B \,\xi^{\mu}_{(\phi)}- \frac{Q-2 a M B}{2 M}\,
\xi^{\mu}_{(t)} \,.
\label{pot11}
\end{equation}
We see that in addition to the usual magnetic and Coulomb parts
of the electromagnetic field the 4-potential (\ref{pot11})
also contains a contribution proportional to the rotation parameter
of the black hole. It is clear that such a field appears due to
the Faraday induction; a rotation of the Kerr metric produces an induced
electric field, just as a field would be induced by a rotating loop
in a magnetic field. The underlying geometry is such that the induced
potential difference arises between the event horizon and infinitely
remote point
\begin{equation}
\Delta \Phi = \Phi_H - \Phi_{\infty}=\frac{Q-2 a M B}{2 M} \;.
\label{potd}
\end{equation}
Since an astrophysical object will rapidly neutralize
its electric charge by a process of the selective accretion of charges
from surrounding plasma, the potential difference
(\ref{potd}) will vanish, or equivalently
\begin{equation}
\tilde Q= Q-2 a M B= 0
\label{charge}
\end{equation}
and the black hole will acquire an inductive electric charge
$ Q=2a M B $ . On the other hand, in this very simple model
we see that the Faraday induction may be a possible mechanism
for the tapping of the rotational energy from a black hole. Namely,
this mechanism lies in the basic structure of the model of a supermassive
black hole for active galactic nuclei and quasars
(Blandford $\&$ Znajek 1977; Kardashev 1995) .
\vspace{4mm}

\section{Motion of charged particles}

In a realistic model of gas accretion onto a black  hole
energy is released mostly at the expense of the binding energy of the test
particles moving in a strong gravitational field of the black hole.
This may be used to propose an alternative model for the interpretation
of the observational data from the core of a number of galaxies
(see Shapiro $\&$ Teukolsky 1983 and references therein).
Moreover, as we have mentioned above the observations of the core of some
galaxies reveal the existence of
non-thermal radiation and radio waves. The latter can be explained
by synchrotron emission from ultrarelativistic electrons in
a strong magnetic field in the innermost orbits
around a supermassive black hole (Rees 1998).
This makes it very important to investigate in detail
the motion of charged particles
around a rotating black hole in an external magnetic field.
In the following we shall do it using (\ref{pot11}) as
a 4-vector potential of the external magnetic field.
For convenience, we shall use the gauge
in which $A^0=0$ at infinity. Then we have
\begin{equation}
\tilde {A^0}= \frac{\tilde Q r (r^2+a^2)}{\Delta \Sigma}\;,\;\;\;\;\;
\tilde {A^{\phi}}= \frac{B}{2} +\frac{\tilde Q r a}{\Delta \Sigma}.
\label{potcom}
\end{equation}
We shall study the motion of the test particles around a rotationg
black hole with zero electric charge $(\tilde Q \rightarrow 0)$
using the Hamilton-Jacobi equation
\begin{equation}
g^{\mu\nu} \left(\frac{\partial S}{\partial x^{\mu}} +e \tilde A_{\mu}\right)
\left(\frac{\partial S}{\partial x^{\nu}} +e \tilde A_{\nu}\right)=m^2 ,
\label{hj}
\end{equation}
where $e$ and $m$ are the charge and the mass of a test particle,
respectively. Since  $t$ and $\phi$ are the Killing variables we
can write the action in the form
\begin{equation}
S=-E t + L \phi + f(r,\theta) ,
\label{action}
\end{equation}
where the conserved quantities $E$ and $L$ are the energy and the angular
momentum of a test particle at infinity. Substituting it into equation
(\ref{hj}) we come to the equation for unseparable part of the action
\begin{eqnarray}
\Delta \left(\frac{\partial f}{\partial r}\right)^2 +
 \left(\frac{\partial f}{\partial \theta}\right)^2 -\frac{A}{\Delta}\,E^2+
 \frac{\Sigma-2 M r}{\Delta \sin^2\theta}\,L^2 +\frac{4 M r a}{\Delta}\, E L
 -e B L \Sigma
\nonumber \\
 +\,\frac{1}{4}\,e^2 B^2 A \sin^2\theta  +m^2 \Sigma=0
\label{eq1}
\end{eqnarray}
It is not possible to separate variables in this equation
in general case, however
one can separate it in the equatorial plane $\theta=\pi/2$. Then we
obtain the equation for radial motion
\begin{equation}
r^3 \left(\frac{d r}{d s}\right)^2=V(E,L,r,\epsilon)
\label{eq2}
\end{equation}
where $s$ is the proper time along the trajectory of a particle and
\begin{eqnarray}
V=(r^3 + a^2 r + 2 M a^2)\, E^2 -(r-2 M)\,L^2 - 4 M a E L
\nonumber \\
 - \Delta r \,(1+\frac{\epsilon\,L}{M})  -\frac{\Delta}{4 M^2}\, \epsilon^2\,
(r^3 +a^2 r + 2 M a^2)
\label{effpot}
\end{eqnarray}
can be thought of as an effective potential of the  radial motion.
Here we have changed $ E\rightarrow E/m $ and $ L\rightarrow L/m $.
The effective potential besides the energy,
the angular momentum and the radius of the motion
also depends on the dimensionless parameter
\begin{equation}
\epsilon= \frac{eBM}{m} \;,
\label{inflpar}
\end{equation}
which characterizes the relative influence of a uniform magnetic field
on the motion of the charged particles. We shall call it as
the influence parameter of the magnetic field.
We note that even for small values of
the magnetic field strength $( B/B_M \ll 1)$  the parameter  $\,\epsilon\, $
for a particle with high charge-to-mass ratio
(for instance, for electron $e/m \simeq 10^{21}$) may not be small.
\vspace{4mm}

\section{Marginally stable circular orbits}

We shall now describe a particular class of orbits, namely
circular orbits that play an important role in
understanding the essential features of the dynamics of test particles
around a rotating black hole in a magnetic field. Physically,
from the symmetry of the problem it is clear that the circular orbits
are possible in the equatorial plane $\theta=\pi/2$ and
to describe them one must set $ {d r \over d s} $ to be zero.
This, in turn, requires vanishing of the effective potential (\ref{effpot})
\begin{equation}
V(E,L,r,\epsilon)=0
\label{eq3}
\end{equation}
along with its first derivative with respect to $r$
\begin{equation}
\frac{\partial V(E,L,r,\epsilon)}{\partial r} =0
\label{eq4}
\end{equation}
The simultaneous solution of these equations would determine
the energy and the angular momentum of the circular motion
in terms of the orbital radius, the hole's rotation parameter and
the influence parameter of the magnetic field $\,\epsilon\,$.
However, the underlying expressions involve high order polynomial
equations and their analytical solution is
formidable and defies analysis. Therefore in
Prasanna \& Vishveshwara (1978) and Prasanna (1980) the authors have
appealed to a numerical analysis of equations
(\ref{eq3}) -(\ref{eq4}) for different
values of the constants of motion, orbital radii, black hole's
rotation parameter, as well as the strength parameter of
the magnetic field. 

Fortunately, the situation
is changed if one wishes to restrict oneself to considering
only the marginally stable circular orbits.
After all, the stable orbits are most of interest
astrophysically, as the binding energy of the marginally stable
circular orbits is of an energy source for the potential efficiency of an
accretion disc around a black hole. The stability of the
circular motion requires the relation
\begin{equation}
\frac{\partial^2 V(E,L,r,\epsilon)}{\partial r^2} \leq 0
\label{eq5}
\end{equation}
where the case of equality corresponds to the marginally stable
circular motion. It is clear that the simultaneous solutions of
equations (\ref{eq3}) -(\ref{eq5}) will determine the region of
stability, the associated energy and angular momentum of
the circular orbits. It is remarkable that in this case one can
obtain the close equation governing the stability region which 
depends only on the rotation parameter of the black hole and
the influence parameter of the magnetic field. Below we shall
show that for the limiting values of the parameters this equation
admits the simple analytical solutions for the radii of stability.

From equations (\ref{eq4}) and (\ref{eq5})
we find that the angular momentum and the energy of a test particle
can be given in the form
\begin{equation}
L = -\epsilon \left(r - {a^2\over {3 r}} \right) \pm \sqrt{\lambda}
\label{angm}
\end{equation}

\begin{equation}
E = \left[\eta \mp {\epsilon\over M}
\left(1-{{2 M}\over {3 r}} \right) \sqrt{\lambda}\right]^{\;1\over2} ,
\label{energy}
\end{equation}
where we have used the notations
\begin{eqnarray}
& &\lambda= 2 M\left(r-{a^2\over 3 r}\right)
+{\epsilon^2 \over 4 M^2}\left[r^2 \left(5 r^2- 4 M r+4 M^2 \right) +
\right. \nonumber  \\ & & \left.
{2\over 3} \,a^2 \left(5 r^2-6 M r+2 M^2 \right)
+a^4 \left(1+{4 M^2\over 9 r^2}\right)\right]
\end{eqnarray}
and
\begin{equation}
\eta= 1- \frac{2 M}{3 r}-{\epsilon^2\over 6}\left[ 4- 5{\,r^2\over M^2}
-{a^2\over M^2}\left(3-{2 M\over r}+{4 M^2 \over 3 r^2}\right)\right]
\end{equation}
These expressions depend on the radius of stability of a circular orbit
which yet has to be determined. Substituting (\ref{angm}) and
(\ref{energy}) into equation (\ref{eq3}) we obtain the
equation
\begin{eqnarray}
& & \left(6 M r - r^2 + 3 a^2 - {4 a^2 M\over r} \right)
\left(1\mp {\epsilon \over M} \sqrt{\lambda}\right)
\nonumber \\[2mm] & &
+ \,\epsilon^2 \left[ r^2\left(6- {4 r\over M} +{9\over 2}{r^2\over M^2}
-{r^3 \over M^3} \right)
\right. \nonumber \\[2mm] & & \left.
-a^2 \left(3 + {8 \over 3}{ r\over M} - {5 r^2\over M^2}\right)
+ {3 \over 2}{a^4 \over M^2} \left(1-{2 M \over 3 r} +{8 M^2\over 9 r^2}
\right) \right]
\nonumber \\[2mm] & &
 \mp \,6 a
\left[\sqrt{\lambda} \mp \epsilon \left(r - {a^2\over {3 r}} \right)
\right] \left[\eta \mp {\epsilon\over M} \left(1-{2\over 3}
{M\over r}\right) \sqrt{\lambda}\right]^{1\over2} = 0 .
\label{radius}
\end{eqnarray}
The solution of this equation will determine the radii of the
marginally stable circular orbits as functions of the
rotation parameter $\, a \,$, as well as
of the influence parameter of the magnetic field $\,\epsilon \,$.

It should be noted that the circular motions will occur in
direct and retrograde orbits depending on whether the particles are
corotating $\, (L >0 \, ) $, or counterrotating $\, (L < 0)\,$ with respect
to the rotation of a black hole. In addition, the presence of
a uniform magnetic field will produce the Larmor and anti-Larmor
motions depending on whether the Lorentz force points
toward a black hole, or it points in the opposite direction.
We recall that we are considering only the case of parallel
alignment of black hole's rotation axis and the direction of a
uniform magnetic field. Therefore, as we shall see below,
the Larmor motion $\, (L < 0)\,$ will occur in retrograde orbits,
while the anti-Larmor motion $\, (L > 0)\,$ will accompany
the direct orbits. Thus, in all expressions above
the upper signs corresponds to the direct,
or the anti-Larmor orbits and the lower signs refer
to the retrograde, or the Larmor orbits. Next, before carrying out
the full analysis of equations (\ref{angm}), (\ref{energy}) and
(\ref{radius}), it is useful to proceed with the particular cases.

\vspace{4mm}

\subsection{Kerr black hole}

In the absence of a magnetic field the energy and the angular momentum
of a test particle in the marginally stable circular orbits around
a Kerr black hole are obtained from equations (\ref{angm}) and
(\ref{energy}) for $\,\epsilon=0 \,$. We have
\begin{equation}
E=\sqrt{1-{2 M \over 3 r}},\;\;\;\;\;\;
L = \pm \,\sqrt{2 M\left(r - {a^2\over {3 r}} \right)},
\label{keangm}
\end{equation}
where, the radii $ r=r_{ms} $ of the orbits satisfy the equation
\begin{eqnarray}
&& 6 M r^2-r^3-4 a^2 M + 3 a^2 r \,\mp
\nonumber \\[2mm] & &
2\sqrt{2}\, a M^{1/2}
\left[\left(3\,r-2M \right)\left(3 r^2-a^2\right)\right]^{\,1\over2}=0
\label{kms}
\end{eqnarray}
which is of a particular case of equation (\ref{radius})
for  $\,\epsilon=0 \,$. Since the radii of stability are different for
direct and retrograde motions, their appropriate energy and angular
momentum given in equations (\ref{keangm}) are different as well.
The solution of (\ref{kms}) can be written in the form
\begin{equation}
r_{ms}=M \left\{3+\sqrt{k_1+k_2} \mp\left[2k_1-k_2
-\frac{16(3-k_1)}{\sqrt{k_1+k_2}}\right]^{1 \over 2} \right\}
\label{rms}
\end{equation}
where
\begin{eqnarray}
k_1 &=& 3+{a^2\over M^2} ,
\nonumber \\
k_2 & = & \left(3-{a\over M}\right)\left(1-{a\over M}\right)^{1\over3}
\left(1+{a\over M}\right)^{2\over3}+
\nonumber \\ & &
\left(3+{a\over M}\right)
\left(1+{a\over M}\right)^{1\over3}\left(1-{a\over M}\right)^{2\over3}
\label{rmk}
\end{eqnarray}
This expression agrees with that found by Bardeen et al. long
ago (Bardeen et al. 1972). For a nonrotating
black hole, $\, a=0 \,$, it gives $ r_{ms}=6 M $, while in the
maximally rotating case, $\,a=M\,$, one finds $\,r=M\,$
for the direct orbits and $\,r=9M\,$ for the retrograde orbits.
Taking these into account in (\ref{keangm})
we find the limiting values for the energy of the direct and retrograde
motions
\begin{equation}
E_{direct}= \frac{1}{\sqrt{3}} \;,\;\;\;\;\;\;
E_{retrograte}= \frac{5}{3\sqrt{3}} .
\label{feangm}
\end{equation}
In order to make the above limits more trasparent and also
with the purpose of comparing them with the corresponding
quantities in the presence of an external magnetic field
Table 1  provides the full list for the values of the radii,
of their assigned energies and angular momenta
as functions of the rotation parameter of a black hole.
It is important to note that the limiting value of
the energy of a direct orbit determines the maximum
binding energy that can be assigned to a last stable circular orbits.
One can easily find that this quantity is of order of $ 42\% $ of
the rest energy and it determines the efficiency of
an accretion disc around a maximally rotating black hole.
\vspace{4mm}

\subsection{Schwarzschild black hole in a magnetic field}

From equations (\ref{angm}) and (\ref{energy}) it is clear that the
case $\, a=0\,$ will give us the angular momentum and
the energy of a charged particle moving in a marginally
stable circular orbit around a nonrotating black hole
immersed into a uniform magnetic field
\begin{equation}
L = -\epsilon \,r \pm {{r^{1/2}}\over {2 M}}\;\Lambda
\label{sangm}
\end{equation}
and
\begin{equation}
E =  \left[\left(1-{{2 M}\over {3 r}} \right)
\left(1\mp \frac{\epsilon \,r^{1/2}}{2 M^2}
\,\Lambda \right)
+ \,\frac{\epsilon^2}{6}\, \left({{5 r^2}\over M^2} -4 \right)
\right]^{1\over2}
\label{senergy}
\end{equation}
where

$$ \Lambda= \left[8 M^3+ \epsilon^2 r\, (5 r^2- 4 M r+4 M^2)
\right]^{1\over2}  $$
and the stability radius $\,r=r_{ms}\,$ is determined by the equation
\begin{equation}
\left(6 M r - r^2 \right)
\left(1 \mp \,{\epsilon\, r^{1/2}\,\over 2 M^2} \,\Lambda \right)
+\,\epsilon^2 r^2 \left(6-{4 r\over M}+{9\over 2}{r^2\over  M^2}
-{r^3\over M^3} \right)=0
\label{srms}
\end{equation}
As we have mentioned above in this case there exist two different
kind of motions depending on the directions of the Lorentz force
with respect to a black hole. In turn, it is
associated with two signs in equations (\ref{sangm})-(\ref{srms}).
Following to papers by Gal'tsov $\&$ Petukhov (1978) and
Aliev $\&$ Gal'tsov (1989) we shall distinguish the two kind of motions
as the Larmor (the lower signs in (\ref{sangm})-(\ref{srms}))
and the anti-Larmor (the upper signs in (\ref{sangm})-(\ref{srms}))
motions. In order to get more insight into these motions it is useful
to obtain the expressions for the angular momentum and energy
in the asymptotically flat region $\,r\gg M\,$.
Solving equations (\ref{eq3}) and (\ref{eq4}) simultaneously
for $\,r\gg M\,$ we find that for orbits with $ L < 0 $
\begin{equation}
L \simeq -{{e B r^2}\over {2 m}}\;,\;\;\;\;\;\;
E \simeq \sqrt{1+
\,\left(\frac{e B r}{m}\right)^2 } .
\label{asexp}
\end{equation}
It follows that this kind of motion is nothing but an ordinary
cyclotron rotation in a uniform magnetic field under the Lorentz force
pointing towards the centre of the orbit of typical radius
\begin{equation}
r=\frac{m v}{e B \sqrt{1-v^2}}
\end{equation}
This is the Larmor motion, while in the opposite case of orbits
with $\,L>0 \,$ we  find that $\,E \rightarrow 1\,$,
that is the anti-Larmor motion when the Lorentz force points
outwards the centre of the motion may occur only in the presence of
a black hole. Next, we need to solve equation (\ref{srms})
for $\, r\,$. It is of a high order polynomial equation and only
for large values of $\,\epsilon \,$
$\,(\epsilon \gg 1) $ one can find the limiting solutions
\begin{equation}
r_{L} = \frac{5+\sqrt{13}}{2} \; M \,,\;\;\;\;\;\; r_{A} \rightarrow 2 M .
\label{limit1}
\end{equation}
Note that the radii of the marginally stable circular orbits
are different for the Larmor $\,(r_{L}\,)$ and anti-Larmor $\,(r_{A}\,)$
motions, the effect of a uniform magnetic field shifts both of them towards
the event horizon and even right up to the event horizon
for the anti-Larmor motion when $\,\epsilon\,$ is large enough.
These conclusions are in agreement with those made in
Gal'tsov $\&$ Petukhov (1978) on the basis of numerical calculations.
Substituting the radii (\ref{limit1}) in equations
(\ref{sangm})-(\ref{senergy}) respectively, we find that for
$\,\epsilon \gg 1\,$
\begin{equation}
E_{L} \rightarrow 5.56 \,\epsilon  \,M \,,\;\;\;\;\;\;
L \rightarrow -23.47 \,\epsilon \,M
\label{limlar}
\end{equation}
for the Larmor motion and
\begin{equation}
E_{A} \rightarrow 0 \,,\;\;\;\;\;\;
L \rightarrow 2 \epsilon \,M
\label{limanti}
\end{equation}
for the anti-Larmor motion.
First of all, it follows from (\ref{limlar})
that for $\,\epsilon \gg 1\,$ there exists a stable ultarelativistic motion
of charged particles, while in the absence of a magnetic field as is well
known, a stable motion in the field of a nonrotating black hole
is only non-relativistic. In addition, the energy of
the anti-Larmor motion in the innermost stable orbit tends to zero
and the corresponding binding energy approaches $\,\simeq 100 \% \,$
of the rest energy, in contrast to $\,42\% \,$ that of in
the field of a maximally rotating black hole.
A more detailed analysis of how the above conclusions are
made is given in Table 2. It exhibits the full list for the values of the
radii and of the associated energies and angular momenta as functions
of the influence parameter of the magnetic field $\,\epsilon \,$.
\vspace{4mm}

\subsection{General case}

We now turn to the consideration of the combined effects of the black hole
rotation and a uniform magnetic field on the marginally stable
circular orbits. The radii of their stability are determined by equation
(\ref{radius}), which in the general case can be solved only numerically.
However, in the case of a maximally rotating black hole and
large enough values of the magnetic field strength $\,(\epsilon \gg 1) \,$
we find the limiting analytical solution for a retrograde motion
\begin{equation}
r = 2 M \left\{1+2 \cos \left[{1\over 3}\,
tan^{-1} \,\left({\sqrt{7} \over {3}}\right) \right] \right\}
\label{limit2}
\end{equation}
while, for a direct motion we have the two different radii
\begin{equation}
r_{1} \rightarrow M \;,\;\;\;\;\;\;\;
r_{2} = (1+\sqrt{2})\,M .
\label{limit22}
\end{equation}
The appearance of the two stable orbits can apparently be
related to the effect of the expelling of magnetic flux lines
from the horizon of a black hole as  the rotation
parameter of the hole increases (Bi\u c\'ak $\&$ Jani\u s 1985).
In our case a large enough value of the magnetic field
shifts the innermost stable
circular orbits to the event horizon and this, in turn,
along with the increase of the hole's rotation parameter
provides the expulsion of the magnetic field lines from the black hole.
As a result of this the above two anti-Larmor orbits appear.
In the case of the first orbits the radius of stability tends
the event horizon, but the effect of the magnetic field on
it should become less and less as it approaches the horizon.
In the second case the anti-Larmor motion is expelled from the black hole
and it occurs at the limiting radius given by $\,r_2\,$ in (\ref{limit22}).

A more precise description of it, of course, requires
the detailed numerical analysis of equation (\ref{radius}). The results
are listed in Tables 3.1-3.2. Comparing these results
and also the radii (\ref{limit2})-(\ref{limit22})
with those given in subsection 4.1  we see that
the magnetic field always plays a stabilizing role in its effect on
the circular orbits. In the Kerr metric the region
of stability for direct orbits enlarges towards the event horizon
with the increase of hole's rotation parameter, while for prograde orbits
it moves in the opposite direction (see Table 1). The presence of
a large-scale uniform magnetic field around a Kerr black hole shifts
the innermost stable circular orbits towards the horizon
for both direct and retrograde motions.

However, as it is seen from Table 3.2,
when $ \,\epsilon \,$ is large enough the
direct motion decays into two motions
and the separation becomes more significant as
the rotation parameter tends to its maximum value.
This is due to the effect of the
expelling of the magnetic field from nearby region of the horizon
as the angular momentum of the black hole increases.
At the same time, Tables 3.1-3.2. show that
for retrograde orbits the rotation of a black hole
opposes to the stabilizing effect of the magnetic field,
leading finally to the limiting value of
the radius given in (\ref{limit2}).

Some results of the numerical analysis of equation (\ref{radius})
are also depicted in Figs.1-2.  Fig.1 shows the
dependence of the radii of marginally stable
circular orbits on the rotation parameter $\, a \,$ for
values of $\,\epsilon = 0, 1 $. In Fig.2  for various values of
the black hole rotation parameter
we illustrate the enlargement of the region of the marginal stability with
the growth of the influence parameter of the magnetic field.
In both cases we observe that even
not very large value of the magnetic field produces
an essential enlargement in the region of the marginally stable
circular orbits towards the horizon of the black hole.

Next, substituting the value of the radius
(\ref{limit2}) into equations (\ref{angm})-(\ref{energy})
we calculate the limiting value for the pertaining
energy and angular momentum at a retrograde motion
\begin{equation}
E = 7.82 \,\epsilon \,,\;\;\;\;\;\;
L = - 42.55 \, \epsilon \,M .
\label{limret}
\end{equation}
As $\,\epsilon \gg 1  \,$, the retrograde motion of a charged particle
at the last stable circular orbit  around a maximally rotating black hole
is of a relativistic motion as in the case of a
nonrotating black hole (see equation (\ref{limlar})), however,
the rotation further enhances this effect. In the same way,
using the radius $\, r_{2} \,$ in (\ref{limit22}) we find the energy and
the angular momentum
\begin{equation}
E = 0.41 \,\epsilon  \,,\;\;\;\;\;\;
L = 3.82 \,\epsilon \,M
\label{limdir}
\end{equation}
pertaining to the innermost direct, anti-Larmor, motion. It follows that
the anti-Larmor motion of a charged particle at the radius $\,r_2 \,$
may also become relativistic for $\,\epsilon \gg 1  \,$,
in contrast to that in the case of a nonrotating black hole,
where the corresponding energy tends to zero (equation (\ref{limanti})).

In Tables 3.1-3.2 we also
give the full list for the values of the energies and angular momenta
pertaining to the marginally stable circular orbits. Comparing these
quantities with those in Table 2  we see that in all cases the energy of
the retrograde motion monotonically increases up to its
relativistic values as the influence of the magnetic field, as well as
the rotation of the black hole increase. However,
the energy of the anti-Larmor motion around a Schwarzschild black hole
systematically decreases as a test particle approaches
the event horizon. The rotation of the black hole
drastically changes the situation. The energies of direct motions
increase monotonically as the hole's
rotation parameter increases, excepting the nearest to the horzon region,
where the effect of the expelling of the magnetic field becomes
dominant (Bi\u c\'ak $\&$ Jani\u s 1985). Thus, we may conclude that
for large enough values of the magnetic field strength,
$\,\epsilon \gg 1  \,$,
both retrograde and direct motions of charged particles
around a rotating black hole
become relativistic along the innermost stable circular orbits. It is
especially important to note the possibility of the relativistic stable
direct orbits near the event horizon of a rotationg black hole
in the presence of a strong magnetic field, unlike the case of
a nonrotating black hole. This is due to
the intimate connection between gravitation and electrodynamics
that arises in the model considered here.

\section{Conclusions}
\vspace{4mm}
\noindent
The main purpose of this paper was to study the
combined effects of the rotation of a Kerr black hole and an
external magnetic field on the marginally stable circular motion
of charged particles. We have presented general equations
governing the energy, the angular momentum and the region of
the marginal stability of circular orbits around a rotating black
hole in a uniform magnetic field. The analytical results obtained
for large enough values of the magnetic field strength and for a
maximum value of the black hole angular momentum, as well as the
numerical analysis performed in general case have shown that in
all cases of the circular orbits the magnetic field essentially
enlarges the region of their marginal stability towards the event
horizon. As for the effect of the rotation of a black hole it has
been found to be different for direct and retrograde motions:
For retrograde motion the rotation opposes
to the magnetic field in its stabilizing effect, though the latter
always remains be dominant and for extreme values of the magnetic field
and rotation parameter there exists a limiting value for the
enlargement of the region of stability towards the event horizon.
In the case of direct motion the presence of a rotation
produces an additional shift of the stable orbits
to the event horizon. However, for large enough values of
the magnetic field, when the innermost stable motion occurs
at a radius that lies very close to the event horizon,
the magnetic field lines are expelled from the black hole
as the hole's rotation parameter increases. Accordingly,
the direct motion occurs in two different orbits; one of them
is also expelled from the black hole, while the other one
being affected less and less by the magnetic field approaches
the event horizon as the rotation becomes maximum.

We have shown that the presence of a strong magnetic field
around a rotating black holes provides the possibility of retrograde
motion in the innermost stable relativistic orbits and the rotation of
the black hole further enhances this effect. It is
very important that, unlike the Schwarzschild case,
the combined effects of a sufficiently large magnetic field
and the rotation of a black hole also result in
relativistic motions in the innermost stable direct orbits that lie
very close to the event horizon. Thus, nearby the event horizon
of a rotating black hole there may exist a source of synchrotron
emission from relativistic charged particles moving in
stable circular orbits. This, of course, may play
an important role both in the searches for black holes
and in making feasible the probes of the metric
in the strong-gravity domain.\\[2mm]
\noindent
{\bf {ACKNOWLEDGMENTS}} \\ [2mm]
\noindent
We thank E. \.In\" on\" u, I. H. Duru, M Horta\c csu for their stimulating
interest in our work and C. Sa\c cl\i o\u glu for reading the manuscript
and valuable comments.  \\[4mm]

\noindent
{\bf {REFERENCES}} \\[4mm]
%\vspace{2mm}
%\noindent
Aliev A. N., Gal'tsov D. V., 1989, Usp. Fiz. Nauk, 157, 129\\ \hspace*{4mm}
(Sov. Phys. Usp., 32, 75, English translation) \\[2mm]
Bardeen J. M., Carter B.,  Hawking S. W., 1973,
Commun. \\ \hspace*{4mm}
Math. Phys., 31, 161  \\[2mm]
Bardeen J. M., Press W. H., Teukolsky S. A., 1972, ApJ, \\ \hspace*{4mm}
178, 347\\[2mm]
Bi\u c\'ak J., Jani\u s V., 1985, MNRAS, 212, 899  \\[2mm]
Bi\u c\'ak J., Ledvinka T., 2000, Nuovo Cimento, B115, 739 \\[2mm]
Blandford R. D., Znajek R. L., 1977, MNRAS, 179, 433 \\  [2mm]
Chamblin A., Emparan R., Gibbons G. W., 1998, Phys. Rev. D, 58,\\
\hspace*{4mm} 084009 \\  [2mm]
Charles P. A., 1999, in Abramowicz M. A., Bj\" ornsson G., Pringle J. E.\\
\hspace*{4mm}
eds, Theory of Black Hole Accretion Discs, Cambridge University Press,
\\ \hspace*{4mm}
Cambridge \\  [2mm]
Fabian A. C., Rees M. J., 1995, MNRAS, 277, L55 \\[2mm]
Frolov V. P., Novikov I. D., 1998, Physics of Black Holes,\\ \hspace*{4mm}
Kluwer Academic Press, Dordrecht \\  [2mm]
Gal'tsov D. V., Petukhov V.I., 1978, Zh. Eksp. Teor. Fiz., 74, 801\\
\hspace*{4mm} (Sov. Phys. JETP, 47, 419, English translation) \\[2mm]
Ginzburg V. L., Ozernoy L. M., 1965, Zh. Eksp. Teor. Fiz., 47, 1030\\
\hspace*{4mm} (Sov. Phys. JETP, 20, 689, English translation)  \\  [2mm]
Horowitz G. T., Teukolsky S. A., 1999, Rev. of Modern Physics, 71, 180
\\  [2mm]
Kardashev N. S., 1995, MNRAS, 276, 515 \\  [2mm]
Mashhoon B., 2001, in Pascual-Sanches J. F., Floria L.,\\ \hspace*{4mm}
San Miguel A., Vicente F. eds, Frames $\&$ Gravitomagnetisms,\\
\hspace*{4mm}
World Scientific, Singapore \\  [2mm]
Menou K., Quataert E., Narayan R., 1999, in Iyer B. R., Bhawal B., \\
\hspace*{4mm}
eds, Black Holes, Gravitational Radiation and the Universe, Kluwer, p 265 \\  [2mm]
Merritt D., Oh S. P., 1997, Astron. J. 113, 1279 \\  [2mm]
Miyoshi M., Moran J., Herrnstein J., Greenhill L., Nakai N., Diamond P., 1995, Nature, 
373, 127 \\  [2mm]
Narayan R., Yi I., 1995, ApJ, 444, 231 \\  [2mm]
Prasanna A. R., Vishveshwara C. V., 1978, Pramana, 11, 359 \\ [2mm]
Prasanna A. R., 1980, Rivista del Nuovo Cimento, 3, N.11, 1   \\ [2mm]
Punsly B., 2001, Black Hole Gravitohydromagnetics, Springer  Verlag \\ [2mm]
Rees M. J., 1982, in Riegler G., Blandford R. D., eds, The Galactic \\
\hspace*{4mm} Center, American Inst. of Phys., New York, p 166 \\  [2mm]
Rees M. J., 1998, in Wald R. M. ed, Black Holes and Relativistic Stars,\\
\hspace*{4mm} Chicago University Press, Chicago \\  [2mm]
Richstone D. O., et al., 1998, Nature, 395, 14 \\  [2mm]
Shapiro S. L., Teukolsky S. A., 1983, Black Holes, White Dwarfs and \\
\hspace*{4mm} Neutron Stars, Wiley, New York \\  [2mm]
Thorne K. S., Price R. H., MacDonald D. A., 1986, Black Holes: \\
\hspace*{4mm} The Membrane Paradigm, Yale University Press, New Haven
\\  [2mm]
Wald R. M., 1974, Phys. Rev. D, 10, 1680 \\  [2mm]
Watson W. D., Wallin B. K., 1994, ApJ, 432, L35  \\  [2mm]
Wagh S. M., Dhurandeur C. V., Dadhich N., 1985, ApJ, 290, 12
\newpage
\vspace{1.5cm}
{\bf {Table 1.}} Marginally stable circular orbits around a Kerr
black hole in the absence of a magnetic field
\vspace{4mm}
\begin{center}
\begin{tabular}{|c|ccc||ccc|}
\hline
$\epsilon=0$&&\hspace{-0.8cm} direct orbits\hspace{-0.8cm} &&&\hspace{-1.8cm} retrograde
orbits\hspace{-1.8cm} &\\
\hline
$a/M$&$r/M$& $E$ & $L/M$ &$r/M$& $E$ & $L/M$\\
\hline
 0.0&\, 6.0\,\,& 0.94&3.46\,\,&\,\,6.0\,\,\,\,&0.94& -3.46\quad\\
 0.2&\, 5.32& 0.93& 3.26&\,6.63& 0.94& -3.64\\
 0.4&\, 4.61& 0.92& 3.03&\,7.25& 0.95& -3.80\\
 0.6&\, 3.82& 0.90& 2.75&\,7.85& 0.95& -3.96\\
 0.8&\, 2.90& 0.87& 2.37&\,8.43& 0.95& -4.10\\
\,\,\,\,0.999&\,\,1.18& 0.65& 1.34&\,8.99& 0.96& -4.24\quad\\
\hline
\end{tabular}
\end{center}
\vspace{20mm}
{\bf {Table 2.}} Marginally stable circular orbits around a
nonrotating black hole in a uniform magnetic field
\vspace{4mm}
\begin{center}
\begin{tabular}{|c|ccc||ccc|}
\hline
$\hspace{-0.2cm}a=0\hspace{-0.2cm}$&&\hspace{-2.2cm} anti-Larmor orbits\hspace{-2.2cm} &&&\hspace{-2.2cm} Larmor orbits\hspace{-2.2cm} &\\
\hline
$\epsilon$&$r/M$& $E$ & $L/M$ &$r/M$& $E$ & $L/M$\\
\hline
 0.0&6.0 \,\qquad& 0.94& 3.46\,& 6.0\,\qquad &\, 0.94&\, -3.46\\
 0.2& 3.98& 0.78& 3.51\,&\, 4.63&\, 1.57&\, -6.32\\
 0.4& 3.35& 0.71& 3.92\,&\, 4.41&\, 2.50& -10.35\\
 0.6& 3.05& 0.65& 4.34\,&\, 4.35&\, 3.52& -14.73\\
 0.8& 2.86& 0.62& 4.74\,&\, 4.33&\, 4.59& -19.27\\
 1.0& 2.74&0.58&5.18\,& \, 4.32&\, 5.67& -23.86\\
 1.4& 2.58&0.53&6.03\,& \, 4.31&\, 7.86& -33.11\\
 2.0& 2.43&0.49&7.19\,& \, 4.30&11.16\qquad& -46.99\\
10.0\,\,& 2.10&0.32&23.16\,&\,  4.30&55.60\qquad&-234.42\,\, \\
100.0\,\,\,& 2.01&0.29&203.01\,\,\,&\,  4.30&556.00\,\,\qquad&-2343.99\,\,\,\, \\
\hline
\end{tabular}
\end{center}
\newpage
{\bf {Table 3.1.}} Marginally stable circular orbits around a Kerr
black hole in a uniform magnetic field with $\,\epsilon=1 \,$.
\vspace{4mm}
\begin{center}
\begin{tabular}{|c|ccc||ccc|}
\hline
$\epsilon=1$&&\hspace{-1.8cm} direct orbits\hspace{-1.8cm} &&&\hspace{-1.8cm} retrograde orbits\hspace{-1.8cm} &\\
\hline
$a/M$&$r/M$& $E$ & $L/M$ &$r/M$& $E$ & $L/M$\\
\hline
 0.0& 2.74& 0.58&5.18\,\,&\,\,4.32&5.67& -23.86\\
 0.2& 2.57& 0.62&4.46\,\,&\,\,4.67& 6.16& -27.55\\
 0.4& 2.38& 0.65&3.76\,\,&\,\,4.99& 6.61& -31.17\\
 0.6& 2.17& 0.69&3.10\,\,&\,\,5.31& 7.06& -35.02\\
 0.8& 1.89& 0.71&2.38\,\,&\,\,5.61& 7.49& -38.84\\
\,\,\,\,\,0.999&1.15& 0.59&1.21\,\,&\,\,5.89& 7.89& -42.58\\
\hline
\end{tabular}
\end{center}
\vspace{20mm}
{\bf {Table 3.2.}} The influence of a sufficiently strong magnetic field
on the marginally stable circular orbits around a Kerr black hole
\vspace{10mm}\\
\begin{tabular}
{|l|p{0.29in}llp{0.24in}ll||p{0.24in}p{0.45in}p{0.59in}|}
\hline
$\hspace{-0.2cm}\epsilon=100\hspace{-0.2cm}$&&&direct&orbits&& &retrograde&&orbits\\
\hline
$a/M$&$r_1/M$&\,\,\,$E_1$ & $L_1/M$\,\,\vline&$r_2/M$& \,\,\,$E_2$ & $L_2/M$ &$r/M$&\hspace{0.21cm}$E$&\hspace{0.21cm} $L/M$\\
\hline
 0.0& 2.01&\,\,0.29&203.01\,\vline&2.01&\,\,\,0.29&203.01&4.30&556.00& -2343.99\\
 0.2& 1.98&\,\,9.97&195.55\,\vline&2.05&\,\,\,9.49&219.31&4.65& 605.31& -2714.25\\
 0.4& 1.93&19.74&185.63\,\vline&2.13& 18.46&246.97&4.98& 652.19& -3090.56\\
 0.6& 1.84&29.09&168.84\,\vline&2.23& 26.44&288.74&5.29& 696.60& -3468.64\\
 0.8& 1.68&37.34&141.35\,\vline&2.35& 33.32&343.57&5.59& 739.89& -3857.50\\
 0.999&1.13& 36.15&\,\,\,73.75\,\vline&2.47& 39.55&405.10&5.88& 782.03& -4255.20\\
\hline
\end{tabular}
\newpage
\begin{figure}[!Ht]
\begin{center}
\vspace*{-4cm}
\hspace*{-3cm}       
\epsfig{file=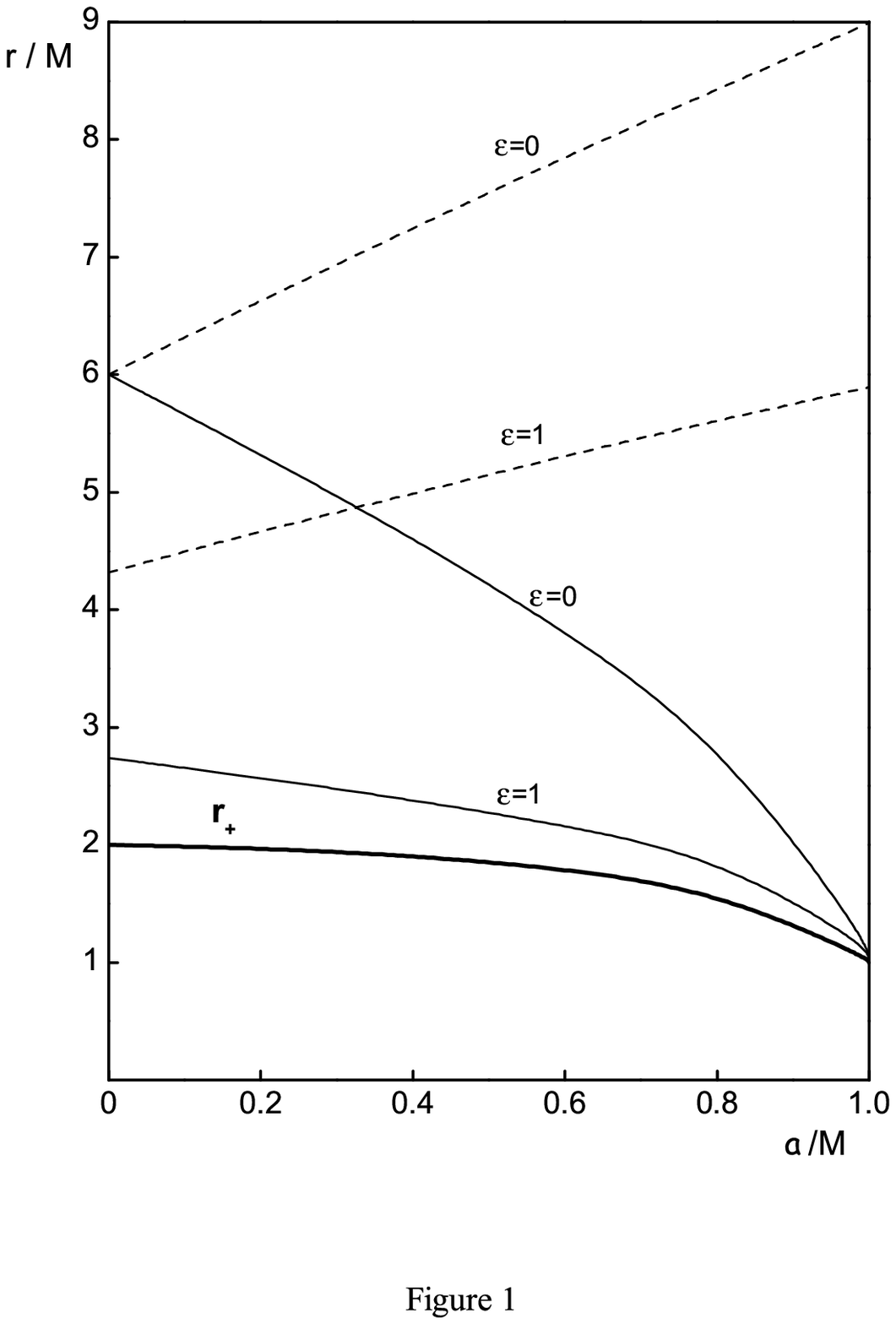,width=18cm,height=25cm}
\vspace*{-5cm}
\end{center}
\end{figure}
\begin{figure}[!Ht]
\begin{center}
\vspace*{-4cm}
\hspace*{-3cm}       
\epsfig{file=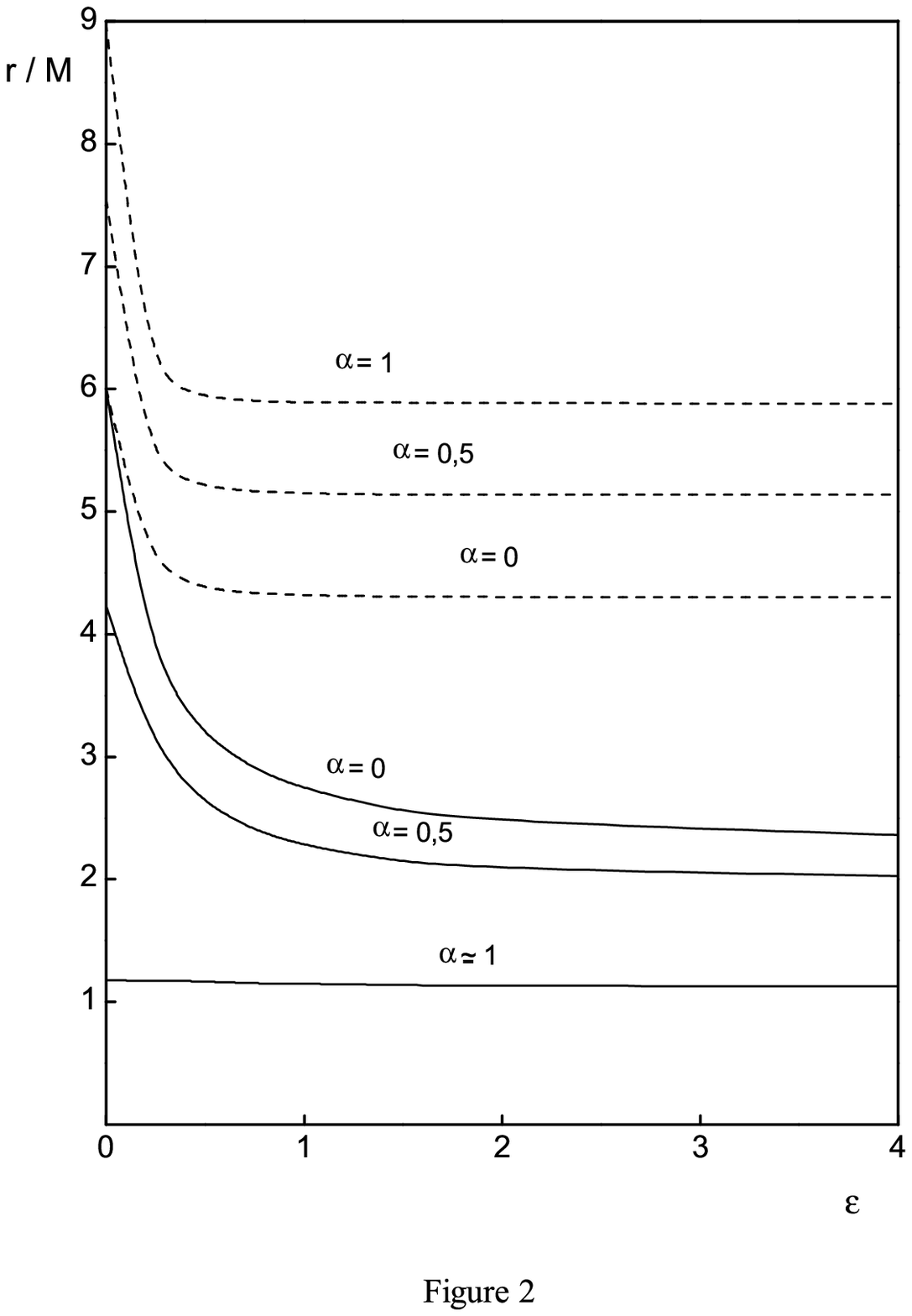,width=17.9cm,height=24.9cm}
\vspace*{-7cm}
\end{center}
\end{figure}

%~\newpage
~\newpage
\noindent
Captions to figures;

\vspace{12mm}
\noindent
{\bf {Figure 1}}. The dependence of the radii of marginally
stable circular orbits around a Kerr black hole on the rotation
parameter $\,a\,$ of the black hole for given $\,\epsilon = 0,\, 1\,$.
The solid curves refer to the innermost direct orbits,
while the dashed curves correspond to the
innermost retrograde orbits, the position of the event horizon is shown
by the bold curve $ \, r_{+}\,$.

\vspace{10mm}
\noindent
{\bf {Figure 2}}. The dependence of the radii of marginally stable
circular orbits around a Kerr black hole in a uniform magnetic field
on the influence parameter of the magnetic field $\,\epsilon\,$.
The solid curves indicate the innermost direct
orbits, while the dashed curves correspond to the innermost
retrograde  orbits $\,(\alpha = a/M = 0,\,  0.5,\, 1)\, $.

\end{document}